\documentstyle[prc,aps,twocolumn,psfig]{revtex}

\def\Journal#1#2#3#4{{#1} {\bf #2}, #3 (#4)}

\def\NPA{{Nucl. Phys.} A}
\def\NPB{{Nucl. Phys.} B}
\def\PLB{{Phys. Lett.}  B}

\def\PRL{Phys. Rev. Lett.}
\def\PRC{{Phys. Rev.} C}
\def\PRD{{Phys. Rev.} D}

\def\ZPC{{Z. Phys.} C}

\def\P{{\mathbf p}}

\def\K{{\mathbf k}}
\def\L{{\mathbf l}}

\def\a{\alpha}

\def\d{\delta}

\def\g{\gamma}

\def\p{\pi}
\def\q{\theta}
\def\m{\mu}
\def\n{\nu}
\def\o{\omega}
\def\s{\sigma}
\def\t{\tau}

\def\cm{{\cal M}}

\def\intkso{\frac{d^3 \K_1}{(2\pi)^3 2\o_1}}
\def\intkst{\frac{d^3 \K_2}{(2\pi)^3 2\o_2}}
\def\intksth{\frac{d^3 \K_3}{(2\pi)^3 2\o_3}}
\def\intlso{\frac{d^3 \L_1}{(2\pi)^3 2 l^0_1}}
\def\intlst{\frac{d^3 \L_2}{(2\pi)^3 2 l^0_2}}

\def\intpsd{\frac{d^3 \P'}{(2\pi)^3 2 E'}}

\def\exp{\mbox{\rm exp}}

\def\lra{\longrightarrow}
\def\llra{\longleftrightarrow}

\def\be{\begin{equation}}
\def\ee{\end{equation}}
\def\bea{\begin{eqnarray}}
\def\eea{\end{eqnarray}}
\def\eref#1{Eq.~(\ref{#1})}

\def\fref#1{Fig.~\ref{#1}}
\def\bfig{\begin{figure}}
\def\efig{\end{figure}}

\newcommand{\ncom}{\newcommand}
\ncom{\vo}[1]{{\mathbf #1}}
\ncom{\vmo}[1]{{|\mathbf #1|}}
\ncom{\vt}[2]{({\mathbf #1}-{\mathbf #2})}
\ncom{\lan}{\langle}
\ncom{\ran}{\rangle}
\ncom\nonum{\nonumber \\}
\ncom\fx{}
\ncom\gsim{\mbox{\raisebox{-0.6ex}{\ $\stackrel {>}{\sim}$\ }}}
\ncom\lsim{\mbox{\raisebox{-0.6ex}{\ $\stackrel {<}{\sim}$\ }}}

\ncom{\half}{{1\over 2}}
\ncom{\third}{{1\over 3}}
\ncom{\fourth}{{1\over 4}}
\ncom{\fifth}{{1\over 5}}
\ncom{\sixth}{{1\over 6}}

\ncom\Tg{T_{eq\; g}}
\ncom\Tq{T_{eq\; q}}
\ncom\qg{\q_g}
\ncom\qq{\q_q}

\begin{document}
\draft

\twocolumn[\hsize\textwidth\columnwidth\hsize\csname @twocolumnfalse\endcsname

\title{Open Charm, Photon and Dilepton Production\\ 
in an \\ Increasingly Strongly Interacting Parton Plasma}
\author{S.M.H. Wong}
\address{Fachbereich Physik, Universit\"at Wuppertal,
D-42097 Wuppertal, Germany}

\maketitle

\begin{abstract}

We examine the effects of the new equilibration scenario of 
the increasingly strongly interacting parton plasma
and of the non-equilibrium environment have on the production
of open charm, photon and dilepton at LHC and at RHIC energies. 
We show that an out-of-equilibrium effect, not shown before,
changes significantly the relative yield of the two main
partonic contributions to photon production in the higher $p_T$ 
range, and higher orders for electromagnetic emissions have 
increased significance in the new scenario especially at RHIC
energies. We argue that the effects of the new scenario
are not restricted just to the parton phase but will 
continue through to the later hadron gas phase with potentially
positive implications for the detection of the quark-gluon plasma.

\end{abstract}

\pacs{PACS number(s): 25.75.-q, 12.38.Mh, 13.87.Ce, 12.38.Bx}
]

\section{Introduction}
\label{sec:intro}

Relativistic heavy ion collision experiments at the present
AGS at Brookhaven and SPS at CERN and at the future colliders
like RHIC and LHC aim to recreate deconfined matter or
the quark-gluon plasma. In order to find out details in
the collisions and to see whether the resulting highly 
compressed matter has made a return trip to the new 
phase through a phase transition, particle productions
in the reactions provide the necessary means for the 
task. However, this is complicated by the fact that
all particle signatures and probes can have their origins
both from hadronic and partonic environment. 
This endeavor would have been much easier if productions 
from deconfined matter were significantly enhanced 
over the hadronic productions. Unfortunately, in practice,
this is not always the case. In fact, one can only hope
for excess in certain finite momentum range of the 
produced particles. The sizes of these windows depend
on details such as the incoming energy/nucleon,
parton distributions in the nucleus, etc.. 
It is vital therefore that production mechanisms behind
both hadronic and partonic origins should be well 
understood and good quantitative control be obtained. 
In this paper, we consider three such particle 
productions from a partonic environment. We study two
electromagnetic and a hadronic probe, namely photons 
and dileptons and open charm. 

Electromagnetic probes are well known to be good probes
because of their much weaker interactions and therefore
they reflect the conditions of the production environment. 
These probes have been studied by many, for example 
\cite{mclerr&toi,kaj&etal,braa&etal,kap&etal,weld,baier&etal1,%
wong4,chak&etal,geig&etal,shury&xion2,trax&etal1,trax&thoma},
in both hadronic and partonic medium which could be in or 
out of equilibrium. Dilepton production could be, for 
instance, used to measure $T_c$ \cite{weld}. Likewise, 
enhancement in soft dilepton production could be due to 
quark-gluon plasma formation \cite{braa&etal,wong4} which 
would provide very good signals. However, background from 
bremsstrahlung off partons and pions \cite{pal&etal} and 
also pion decays \cite{weld} and emission from hadronic 
scattering \cite{baier&etal2} must also be considered.
In the end, one has little choice but to rely on excess in a 
certain restricted momentum range of the produced 
electromagnetic radiation. Something similar can be said 
for photons. Their usefulness in helping to find the equation 
of state \cite{cley&etal}, reveal transverse expansion
\cite{mull&etal} or to see if a phase transition 
has occurred or not \cite{chak&etal} have all been suggested.
In this paper, we do not go into all these other possibilities of
using electromagnetic emissions from the plasma, but
we examine only the emission itself. We look for 
changes if any in the emission rates in the parton phase 
due to a change in the production environment brought about by a 
so-far-neglected new effect and also by the plasma
being out-of-equilibrium. The cause and origin of 
the new effect will be discussed below. 

Open charm is a hard hadronic probe of early dynamics 
\cite{muell&wang,geig2,lin&gyul,lev&etal} whose usefulness 
as a probe depends on the relative yield from the initial 
A+A collision and from the subsequent parton collisions, 
so any effects that could potentially shift the weight from 
one to the other production must be considered. That 
clearly includes parton distributions 
in a nucleus and hence nuclear shadowing \cite{eskola}
effect on the parton distributions and that of gluon in 
particular because of the initial gluon dominance. 
Recent DELPHI experiment at LEP has demonstrated, by
measuring $R^{bl}_3$, the ratio of b-quark to lepton 
3-jet fraction from Z decays, that the b-quark mass 
did run with scale \cite{bern&etal,marti&etal,bil&etal}. 
The running mass $m_b(\m)$ dropped to about 2.67 GeV at the 
Z mass scale. It seems logical therefore that one should 
also take into account the running of the charm quark 
mass when considering open charm production. How important
is this effect for the charm mass will need to be
determined. We will leave this as future investigation. 
In the following, we would not consider the initial 
production which has been done in the previous studies 
of open charm, but rather concentrate entirely in the 
production via the parton plasma. 

Our main motivation in this investigation arises from a 
recently reported new equilibration scenario \cite{wong3} 
which starts right in the middle of that of the 
``hot glue'' \cite{shury} and progressively gains
importance in time. This is the scenario of the Increasingly 
Strongly Interacting Parton Plasma (ISIPP). The interactions 
get stronger and stronger with time in the plasma of quarks 
and gluons because the average parton energies drop due 
to longitudinal cooling and the ``energy sharing'' from 
parton creation. The resulting momentum transfers in the 
parton collisions are bound to decrease as a result. 
Therefore by choosing the most suitable renormalization
scale in the strong coupling to reduce large logarithms 
from higher orders at all time during the time evolution 
of the plasma results naturally with a coupling that 
is increasing in strength with time. 
Note that there are two factors contributing to the decrease 
in momentum transfer, so one cannot hope to get rid of this 
increase in interaction strength in ISIPP just by shutting 
the plasma in a box to stop the expansion and hence 
longitudinal cooling. Parton chemical equilibration 
will make sure that ISIPP is here to stay even if 
there is no expansion. Thus a parton plasma is a rather unique
kind of many-body system.

The effects of the increasing coupling on equilibration have 
been shown in \cite{wong3} using the time evolution scheme 
developed in \cite{wong1,wong2}. This 
new equilibration scenario means that the environment
for particle production, at least, in the later parton 
plasma phase is no longer the same as has been considered
so far. From what we have already mentioned at the beginning,
the effects of this on particle production must be determined 
if one still wishes eventually to identify deconfined 
matter from the particle probes and signatures. 

Another reason for our investigation in particle production
is the environment itself. Because a 
non-equilibrium many-body system is non-trivial and is
tied by its very nature to time evolution, most calculations
restricted themselves to a simplified situation of a
thermalized system. Thus any non-equilibrium effects on particle
productions could not be revealed. We will show the 
presence of this in the case of photon production.

The effects of the ISIPP scenario on particle production
can be roughly divided into two categories. They are
the direct and indirect effects. Direct effect comes 
from those production mechanisms in which strong 
interactions play a part and therefore directly 
depends on $\a_s$. Having said that, this effect is only
operational if the production scale is roughly also the
scale for equilibration. It is not operational if the
scale of the production is always hard as in open
charm production. Indirect effects, as can be guessed
already, have not a direct dependent on the coupling. 
The influence of $\a_s$  comes via its effects on 
the equilibration or time evolution itself. 
More explicitly, they are the effects on the parton 
densities, duration of the parton phase etc.. 
The latter has an important role to play because the
detectors measure what fall into them, when the 
particles were produced is of no consequence,
so it is essential to integrate over 
the history of the collisions before drawing any 
conclusions on the relative yield from hadronic and 
deconfined matter \cite{kap&etal,chak&etal}.      

In the following sections, we will calculate and show
the transverse momentum, $p_T$, or invariant mass, $M$, 
distribution for the three types of particle production
already mentioned. We will compare the productions from a 
parton plasma time evolved with $\a_s=0.3$ with those
from the more consistent ISIPP for which the coupling is
denoted by $\a_s=\a_s^v$ and its value as a function of time
was extracted from the time evolution in \cite{wong3}. 
In the original derivation of the time evolution equations,
they were taken to be centered around the central region
at $z = 0$ or $\eta =0$ where the distributions were
assumed to be more or less uniform. As a consequence, 
we do not perform the integration over spatial rapidity,
or if one prefers, one can multiply by roughly a unit 
rapidity interval assuming uniform distribution around
the central region. So our spacetime integration for 
particle production can be taken to be 
$\int d^4 x = \p R_A^2 \int \t d\t$. We will plot in
all cases, $p_T$ or $M$ distribution at around zero
spatial and particle rapidity. Since we are 
interested in the difference of the standard scenario 
and ISIPP, this should be sufficient to show the 
representative effects of the new time evolution on 
particle productions. 

Our paper is organized as follows. We give the production 
rates and present results of the three types of particle 
production both from a parton plasma time evolved with 
a fixed coupling at essentially the standard value 
$\a_s=0.3$ and from the ISIPP in Sec. \ref{sec:charm}, 
\ref{sec:photon} and \ref{sec:dilepton}. The results
are discussed one by one. Then we move on to a discussion 
on higher orders, implications of ISIPP on particle
production in general beyond the deconfined phase and
other effects of ISIPP in relativisitc heavy ion collisions.

\section{Open Charm Production}
\label{sec:charm}

Open charm production comes from gluon conversions, 
$gg \llra c\bar c$, and quark-antiquark annihilations
$q\bar q \llra c\bar c$. Our results are calculated from 
\bea E {{d^7 N} \over {d^3 p d^4 x}} \fx &=& \fx \frac{1}{2(2\p)^3}
     \int \intkso \intkst \intpsd                            \nonum
     \fx & & \fx \times \; (2\p)^4 \d^{(4)} (k_1+k_2-p'-p)   \nonum
     \fx & & \fx \times
   \bigg \{ \half
    \n_g^2 f_g (k_1,\t) f_g (k_2,\t) |\cm_{gg\lra c\bar c}|^2 
                                                             \nonum  
     \fx & & \fx \;\;\;
   +\n_q^2 f_q (k_1,\t) f_q (k_2,\t) |\cm_{q\bar q\lra c\bar c}|^2
   \bigg \}
\eea
where $\n_g = 2\times 8=16$ and $\n_q=2\times 3\times n_f=6 n_f$
are the multiplicities of gluons and quarks, respectively.  
The distribution functions $f_g$ and $f_q$ are from our previous
investigation on equilibration \cite{wong3,wong2}. 
The matrix elements for $c\bar c$ production are \cite{comb}
\bea |\cm_{gg\lra c\bar c} |^2 \fx &=& \fx \p^2 \a_s^2 
     \bigg \{ \frac{12}{s^2} (m_c^2-t)(m_c^2-u)          \nonum
     \fx & & \fx
             +\frac{8}{3} \left ( \frac{m_c^2-u}{m_c^2-t}
                                 +\frac{m_c^2-t}{m_c^2-u}  
                          \right )                       \nonum 
     \fx & & \fx
             -\frac{16 m_c^2}{3} 
              \left ( \frac{m_c^2+t}{(m_c^2-t)^2}
                     +\frac{m_c^2+u}{(m_c^2-u)^2} 
              \right )                                   \nonum
     \fx & & \fx
             -\frac{6}{s} \Big ( 2m_c^2-t-u \Big )       \nonum 
     \fx & & \fx 
             +\frac{6}{s} \frac{m_c^2 (t-u)^2}{(m_c^2-t)(m_c^2-u)}
                                                         \nonum
     \fx & & \fx
             -\frac{2}{3} \frac{m_c^2 (s-4 m_c^2)}{(m_c^2-t)(m_c^2-u)}
     \bigg \} 
\eea
for gluon conversion into charm-anticharm and
\bea  |\cm_{q\bar q\lra c\bar c} |^2 \fx &=& \fx 
     \frac{64 \p^2 \a_s^2}{9 s^2} 
     \left  \{ (m_c^2-t)^2 + (m_c^2-u)^2 +2 m_c^2 s \right \} \nonum
\eea
for quark-antiquark annihilation. We use an average like 
$\a_s =0.3$ here in the production for both ISIPP and standard 
plasma because as we have already mentioned in Sec. \ref{sec:intro}, 
open charm production is at a hard scale so the increasing coupling
effect for equilibration only affects this through
indirect effect of changes in parton densities and reduced
production time. Since gluon conversion is the main 
contribution and the gluon density in ISIPP is reduced in
general when compared to the standard plasma \cite{wong3},
combining this with reduced production time, there seems
to be an unavoidable significant reduction in open charm
yield. 

In \cite{lin&gyul,lev&etal}, it was shown that charm 
production from initial gluon conversion dominated 
pre-equilibrium production due to the suppression 
coming from spatial and momentum rapidity correlation 
of the produced minijet gluons, and thermal charm 
production was even smaller. Although initial production
is dominant, both \cite{lin&gyul,lev&etal} rely on minijet 
gluon produced from HIJING \cite{gyu1,gyu2,gyu3} for the 
pre-equilibrium production. As remarked in \cite{lev&etal}, 
uncertainties in HIJING warrant some variations of the initial 
conditions at the time $\t_{iso}$ in order to see what these 
uncertainties will lead. If a factor of 4 is given
to the initial parton densities, pre-equilibrium production
can be approximately equal or even larger than the initial 
production by gluon conversion. Thus one can make use of
the sensitivity of open charm production to initial parton
density as a means to probe the latter in high energy
nuclear collisions. In view of the apparent reduction in
yields pointed out above due to the effects of the coupling,
the use of charm as a probe becomes more doubtful. 
This is in fact not the case. The reason will be given
below after we have shown our results of the actual
production and compared the different scenarios.

\bfig
\centerline{
\hbox{
{\psfig{figure=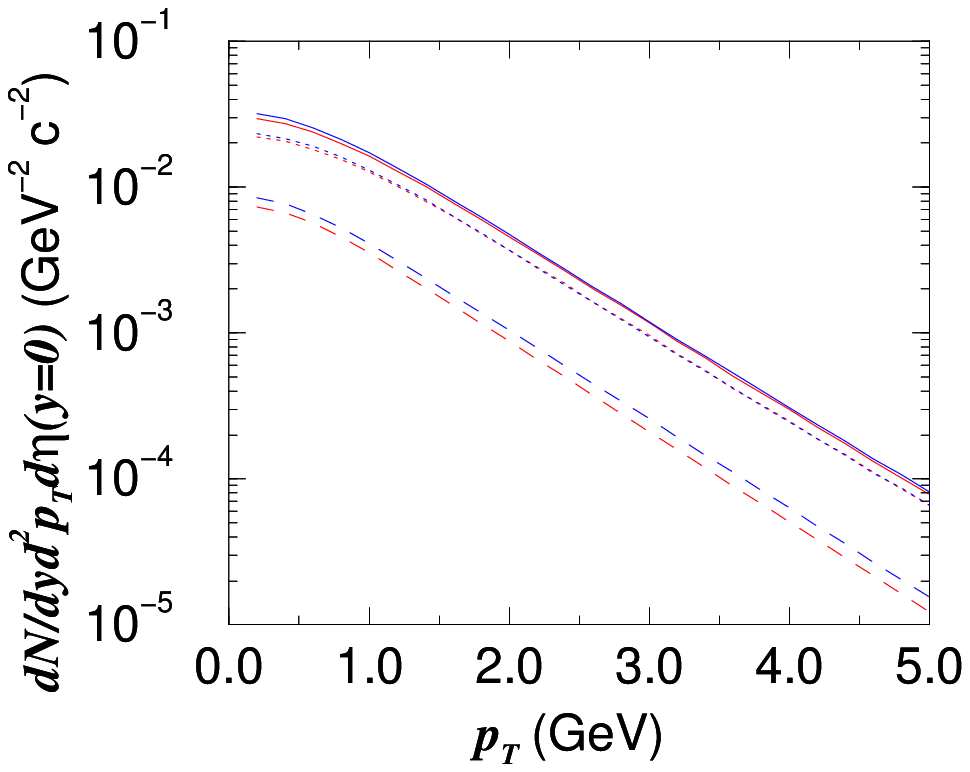,width=2.5in}}
}}
\caption{Comparison of charm production in a parton plasma 
produced at LHC energies, which is time evolved with an evolving 
coupling $\a_s^v$, with the one time evolved with a 
fixed $\a_s=0.3$. For the production itself, we used 
$\a_s=0.3$ in all cases because this is a hard process. 
Dotted and dashed lines are for gluon conversion 
and quark-antiquark annihilation contribution, respectively. 
Solid lines are the sum total. Lines from ISIPP tend to 
lie slightly below the corresponding lines from the fixed
$\a_s =0.3$ evolved plasma. It is more clearly so for 
annihilation contribution.}
\label{f:charm}
\efig

\bfig
\centerline{
\hbox{
{\psfig{figure=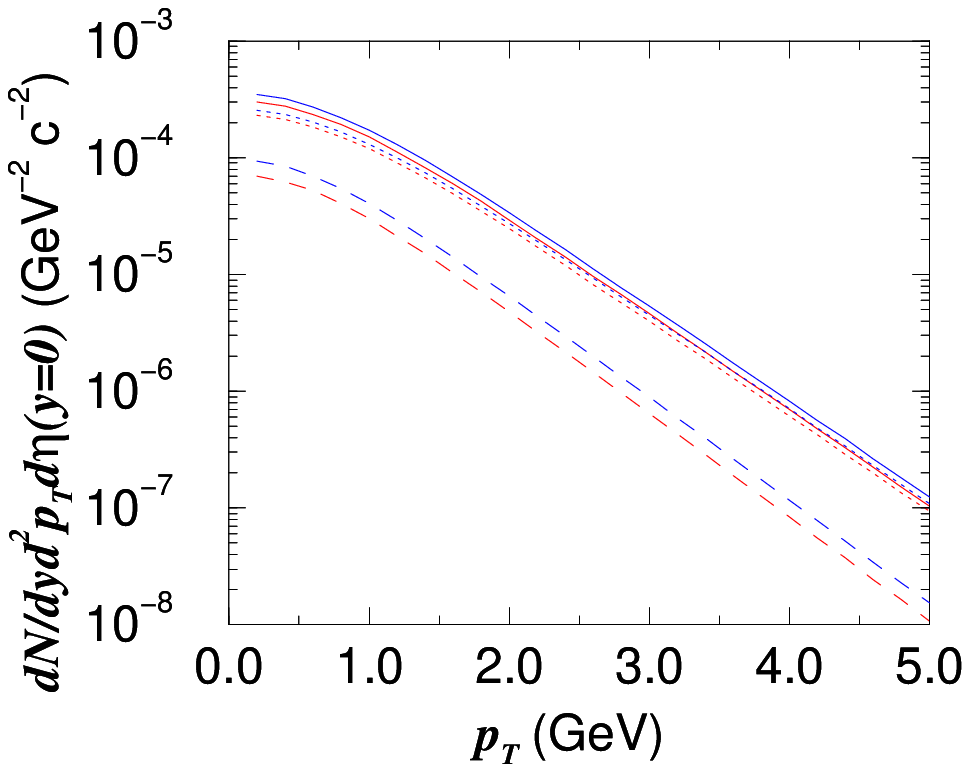,width=2.5in}}
}}
\caption{Same as in \fref{f:charm} but for a parton plasma 
produced at RHIC energies.}
\label{f:charm2}
\efig

In \fref{f:charm} and \fref{f:charm2}, we plotted the charm
yield at LHC and at RHIC as a function of $p_T$ at central 
rapidity. We have not used K-factor in obtaining our plots
because the production ratio of the two scenarios was our
main concern. The dotted and dashed lines are for contributions 
from gluon conversion and quark-antiquark annihilation,
respectively. The full lines are for the total contributions. 
The set of three lines for ISIPP lies just below the corresponding 
lines for $\a_s=0.3$ case. As can be seen from \fref{f:charm} 
at LHC and \fref{f:charm2} at RHIC, both gluon conversion and 
quark-antiquark annihilation contribution from ISIPP 
are slightly reduced when compared with those of the standard 
parton plasma. The reduction of the latter contribution is more 
substantial because although the fermion densities in ISIPP
are higher due to more significant conversion of gluons
into light fermion pairs, this contribution is more spread
out in time, whereas the bulk of the gluon conversion 
contribution tends to be from early times, this we have
checked, and so is less affected by the reduction in the production 
time. For the same reason, the effect of lowered gluon density,
due to the near saturation and the conversion into fermion 
pairs which eventually take toll on the number of
gluons present, essentially does not manifest on the plot. 

The total charm yield at LHC is then about the same in the
two cases and at RHIC, there is a slight reduction.
So although there is no direct effect to help the production, 
due to the different scales of the production itself and that
of equilibration, and there is a reduction in the production time 
which hints at a potential reduction in open charm yield, 
when all factors were considered, the yield is essentially 
unchanged in the parton plasma in the new scenario.
The usefulness of charm as a probe of initial parton densities
concluded in \cite{lin&gyul,lev&etal} is therefore retained.
It is interesting, however, to see how the spread of the
contributions to charm production during the history
of the plasma can avoid certain effects of the coupling
on the final yields.

Having said that a definite conclusion on how good charm as
a probe is cannot be completely settled at present due to 
uncertainties in the nuclear shadowing effect
on the nuclear gluon distribution used to calculate the
charm production from initial gluon conversion. But there
are some recent advances on predicting shadowing effect, 
see \cite{huang&etal}. Also, as already mentioned in the 
Sec. \ref{sec:intro}, the importance of the running mass 
effect must be checked. Clearly all enhancement and suppression 
factors in the production during the two different stages 
must be identified. Here we checked that in the new scenario, 
there is no significant modification to the $p_T$-distribution 
of production from the parton plasma.

\section{Photon Production}
\label{sec:photon}

Photon production comes from Compton scattering $qg\lra q\g$
or $\bar qg\lra \bar q\g$ and quark-antiquark annihilation
$q\bar q \lra g\g$. The production rate is given by
\bea E {{d^7 N} \over {d^3 p d^4 x}} \fx &=& \fx \frac{1}{2(2\p)^3}
     \int \intkso \intkst \intksth                          \nonum
     \fx & \times & \fx \; (2\p)^4 \d^{(4)} (k_1+k_2-k_3-p) \nonum
     \fx & \times & \fx 
   \bigg \{ 2 f_g(k_1,\t) f_q(k_2,\t) (1-f_q(k_3,\t))       \nonum
     \fx & & \fx \;\;\; \times |\cm_{gq\lra q \g}|^2        \nonum
     \fx & & \fx \;  
          + f_q(k_1,\t) f_{\bar q}(k_2,\t) (1+f_g(k_3,\t))  \nonum
     \fx & & \fx \;\;\; \times |\cm_{q\bar q\lra g \g}|^2                
   \bigg \}  \; .
\label{eq:photon}
\eea
A factor of two has been included for the two possibilities
of Compton scattering off quark and antiquark.
The matrix elements, which include colour and spin as well
as infrared screened by the time-dependent medium quark 
mass \cite{wong1,wong2}, 
\be m_q^2 (\t)= 4\p \a_s \; \Big ({{N_c^2-1} \over {2\, N_c}}\Big )
    \int {{d^3\K} \over {(2\p)^3 k}}
    \big (f_g(k,\t) +f_q (k,\t) \big )  \; ,
\ee
are  
\be |\cm_{gq\lra q \g}|^2      = -\sum_q e_q^2 2^9 \p^2 \a \a_s 
    \bigg \{ \frac{s}{t-m_q^2}+\frac{t}{s+m_q^2} \bigg \} \; ,
\label{eq:phot_comp}
\ee
for emission through Compton scattering
and
\be |\cm_{q\bar q\lra g \g}|^2 = \;\sum_q e_q^2 2^9 \p^2 \a \a_s 
    \bigg \{ \frac{u}{t-m_q^2}+\frac{t}{u-m_q^2} \bigg \} \; ,
\label{eq:phot_annih}
\ee
for that through quark-antiquark annihilation. 
In the modulus squared of the matrix elements, we set 
the renormalized coupling $\a_s=\a_s^v$ for the ISIPP
scenario because the scale of the production processes is 
on the average, unlike the hard process of open charm 
production, roughly the same as the scale for parton 
collisions in equilibration. The same also applies to 
dilepton production, which we will consider
up to next-to-leading order in the renormalized coupling 
in Sect. \ref{sec:dilepton} later on. 

The $p_T$ distribution is plotted in \fref{f:phot} and 
\fref{f:phot2} for photon production at LHC and at RHIC,
respectively. The dotted lines are for production
from Compton scattering and dashed lines from
quark-antiquark annihilation. The total are the solid lines.
We have already mentioned the indirect effects of the
coupling on the plasma in the previous section. In photon
production, there is a single power of $\a_s$ in the
modulus squared amplitude in \eref{eq:phot_comp} and
\eref{eq:phot_annih}, which constitutes a direct 
effect on the production. However, in spite of  
the fact that ISIPP evolution caused a drop in the gluon 
density and a reduction in the parton phase duration
in comparison to those of the evolution of the conventional 
plasma, the increase in quark and antiquark densities
plus the progressively increasing interaction strength
compensate for the negative effects in the total sum. 
The curves in the two scenarios lie almost on top of each 
other as a result. 

At this point, we would like to point out an 
out-of-equilibrium effect, which as far as we are aware,
has never been shown before. It is certainly of interest
from a theoretical point of view. We will discuss
its practical significance at the end of this section
once we have shown and explained its origin.

\bfig
\centerline{
\hbox{
{\psfig{figure=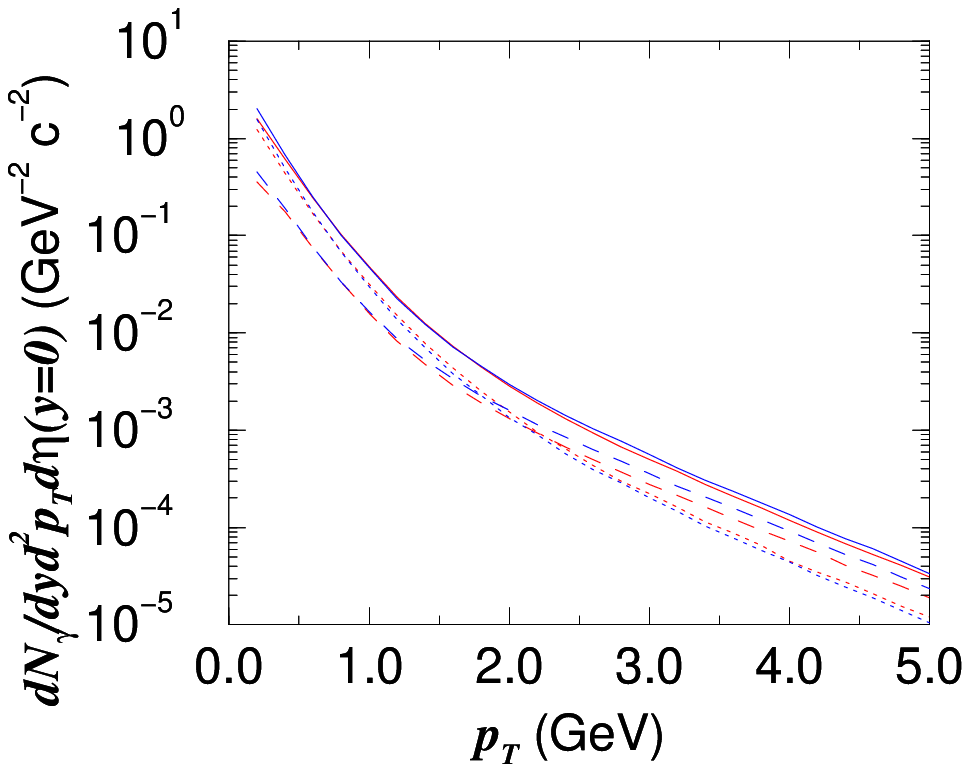,width=2.5in}}
}}
\caption{Photon production from the parton plasma at LHC.
The solid lines are the total sum of the emission from Compton
scattering (dotted) and quark-antiquark annihilation 
(dashed). At large $p_T$, quark-antiquark annihilation 
is the dominant contribution because of the fact that
quarks and gluons are not in equilibrium with respect to 
each other and are therefore at different effective 
temperatures and of quantum statistical effect to a
lesser extent. This contribution from the standard parton 
plasma is slightly above that from ISIPP at large $p_T$.}
\label{f:phot}
\efig

\bfig
\centerline{
\hbox{
{\psfig{figure=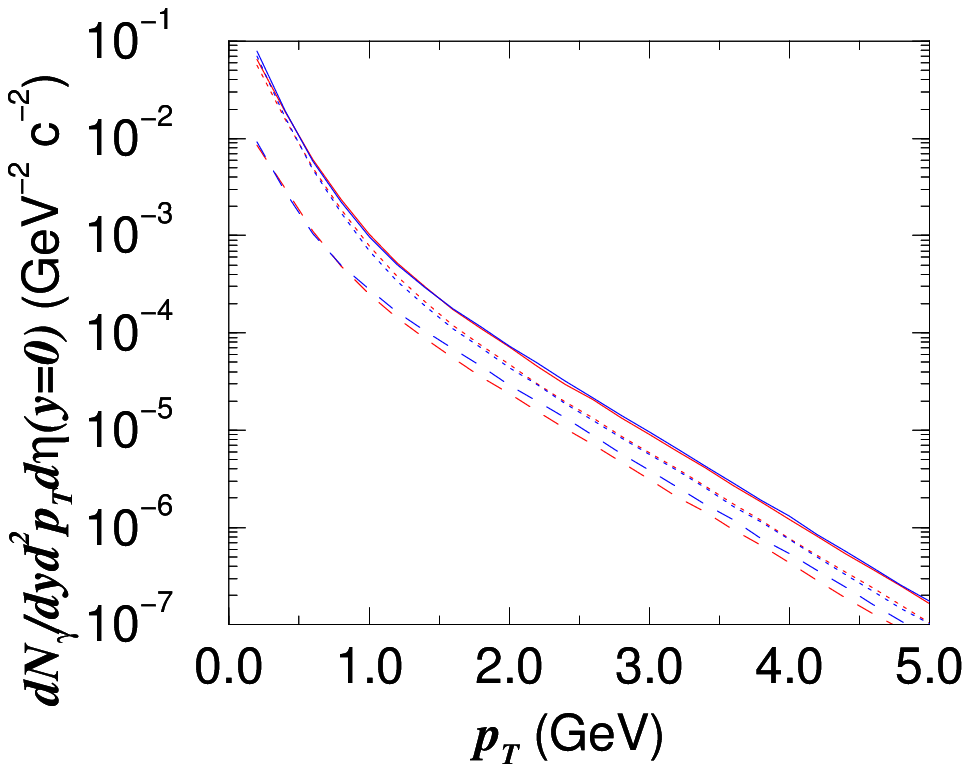,width=2.5in}}
}}
\caption{Same as \fref{f:phot} but at RHIC. In this case, however,
emissions from Compton scattering remain above those from 
quark-antiquark annihilation up to $p_T=$5.0 GeV because of the 
much lower quark to gluon density ratio at RHIC. The point 
where the emission from the latter begin to dominate over the 
former is at higher $p_T$ beyond 5.0 GeV.}
\label{f:phot2}
\efig

In \fref{f:phot}, it is seen that emission from Compton 
scattering at LHC does not dominate over that from 
quark-antiquark annihilation throughout the whole $p_T$ 
range unlike that shown in \cite{mull&etal}. This is 
because in \cite{mull&etal}, the distribution for the 
final state emitted parton has been dropped
and hence quantum statistical effect, such as stimulated
emission, was excluded. The more important reason, however,
is the parton plasma in \cite{mull&etal} was taken 
to be in kinetic equilibrium, an out-of-equilibrium effect 
coming in through the particle distributions of the incoming 
partons could not therefore manifest and, as we will explain 
in the next paragraphs, is the cause of the dominance of 
annihilation over Compton scattering contribution at higher 
$p_T$ at LHC. 

In \cite{wong3,wong2}, we showed that quarks and 
gluons in the plasma were not in equilibrium with 
respect to each other and that they could be considered 
to be at different effective temperatures. The more 
rapid cooling of gluons due to the combined effect of gluon
multiplication and conversion into quark-antiquark
meant that gluons were at a lower temperature effectively most of 
the time than that of quarks and antiquarks. Apart from
the more obvious Pauli blocking, stimulated emission and 
the different matrix elements, the difference between Compton
scattering and annihilation contribution comes from
the distributions $f_g(k_1,\t)$ and $f_q(k_1,\t)$ in 
\eref{eq:photon}. It is this last difference which is
responsible for the effect of the dominance of photon emission 
from quark-antiquark annihilation over Compton scattering
at larger $p_T$.

As can be seen in \fref{f:phot}, the effect is stronger at
higher $p_T$ so we can try to explain it by 
concentrating in this $p_T$ range. Although the plasma is 
not in equilibrium, nevertheless, we can simplify the argument 
by taking $f_g$ and $f_q$ as essentially of equilibrium form 
but at different temperatures. Furthermore, large $p_T$ photon
emission requires high energy incoming partons so
$f_g(k_1)$ and $f_q(k_1)$ can be taken to be of Boltzmann
form $l_g \, \exp(-k_1^0/T_g)$ and $l_q \, \exp(-k_1^0/T_q)$. 
So if the values of the temperatures and fugacities are 
such that $(l_q/l_g) \exp \{k^0_1 (1/T_g - 1/T_q)\} > 1$,
the annihilation contribution will acquire an enhancement
over Compton scattering contribution. This relative
enhancement requires some time to build up as the
effective temperatures have to cool sufficiently so that 
the exponential can more than compensate for the ratio $l_q/l_g < 1$. 
At LHC, the larger ratio of $n_q/n_g$ and the longer duration 
of the parton phase allow the manifestation of this enhancement 
at about $p_T = 1.6$ GeV already. At RHIC, the lower ratio
of $n_q/n_g$ and the shorter duration do not permit
this at $p_T < 5.0$ GeV but the effect is there as one
moves to higher and higher $p_T$ photon. As can be
seen in \fref{f:phot2}, the annihilation contribution
is approaching that from Compton scattering as $p_T$
increases. At low $p_T$, the effect is still there
because both low and high energy incoming partons
contribute but the former are the dominant contribution
in this case, so high $p_T$ photons are needed to select
out harder incoming partons to see this effect. Note that
this is not a density effect since the ratio of the quark
to gluon density $n_q/n_g <1$ always both at LHC 
and at RHIC, but is that of different components 
of the plasma at different effective temperatures. 
The density or fugacity ratio only determines at which
point during the time evolution this effect comes in.
So, by using emission from Compton scattering as a bench
mark, with the out-of-equilibrium and quantum statistical 
effect, photon emission from the quark-gluon plasma can 
have a better chance to compete with direct photon and 
photon fragmented off minijets at large $p_T$ \cite{gupta}
and hence enlarges the window for observing photon emission 
from deconfined matter.

\section{Dilepton Production}
\label{sec:dilepton}

The leading dilepton production rate is, for massless quarks 
and leptons, and adopted to our time-evolving environment,
given by \cite{mclerr&toi,kaj&etal}
\bea \frac{d^8 N^{(1)}_{\m^+\m^-}}{d^4 x d^4 q} \fx &=& \fx 
     \int \frac{d^3 \K_1}{(2\p)^3} \frac{d^3 \K_1}{(2\p)^3} 
     f_q(k_1,\t) f_q(k_2,\t)                              \nonum
     \fx & & \fx \times \d^{(4)}(k_1+k_2-q) 
     v_{rel} \s_{q\bar q \lra \m^+ \m^-} (M)   
\eea
where \ $q^2 = M^2$ is the invariant mass squared of the 
dilepton pair\footnote{We write $\m^+ \m^-$ here to represent
dilepton pair.}, 
$\s_{q\bar q \lra \m^+ \m^-}(M)=12 \sum_q e_q^2 \tilde \s (M)$
is the cross-section for quark-antiquark annihilation into
a dilepton pair. The relative velocity $v_{rel}={M^2}/{2\o_1\o_2}$ 
is that of the quark-antiquark pair and 
$\tilde \s(M) = {4\p \a^2}/{3M^2}$ is the cross-section of
$e^+ e^- \lra \m^+\m^-$. 

\bfig
\centerline{
\hbox{
{\psfig{figure=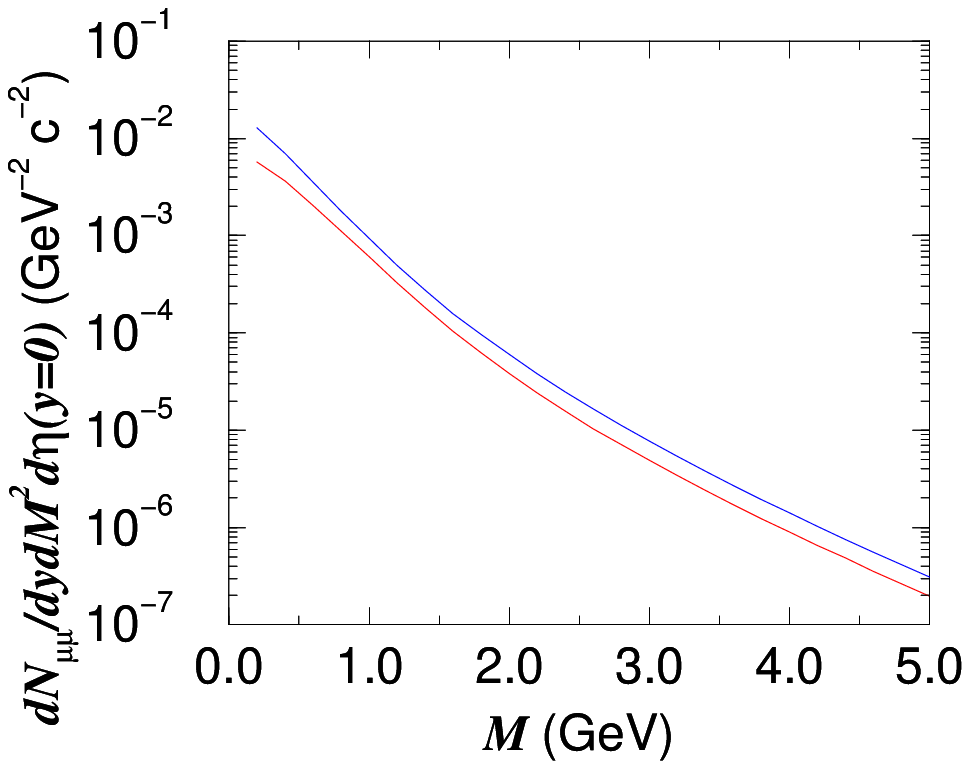,width=2.5in}}
}}
\caption{Comparison of dilepton emission from an ordinary
parton plasma (upper solid line) with ISIPP (lower solid line) 
at LHC. Even though the fermion densities are enhanced in 
ISIPP, the shortening of the duration of the plasma in the 
parton phase is the more important of the two effects. 
So the emission from ISIPP is reduced.}
\label{f:dilep}
\efig

For both standard plasma and ISIPP, the invariant mass 
distributions for dilepton production are shown in 
\fref{f:dilep} and \fref{f:dilep2}. Both at LHC in \fref{f:dilep},
and at RHIC in \fref{f:dilep2}, the dilepton yields from 
ISIPP are below those from the standard parton plasma.
In the absence of direct effect, the enhanced 
quark-antiquark densities are not sufficient
to compensate for the shortened duration of the ISIPP.
The production in ISIPP are down by an approximate factor 
of 1.6 at LHC and 1.9 at RHIC on the average. The 
size of the window for observing dilepton production from
the parton plasma will therefore, unfortunately, be reduced
at least from the production during the time interval
we considered. Of course, to get the complete picture,
one has to take into account production before our
initial time $\t_0$ and from the mixed phase in the 
case of a first order phase transition. 

\bfig
\centerline{
\hbox{
{\psfig{figure=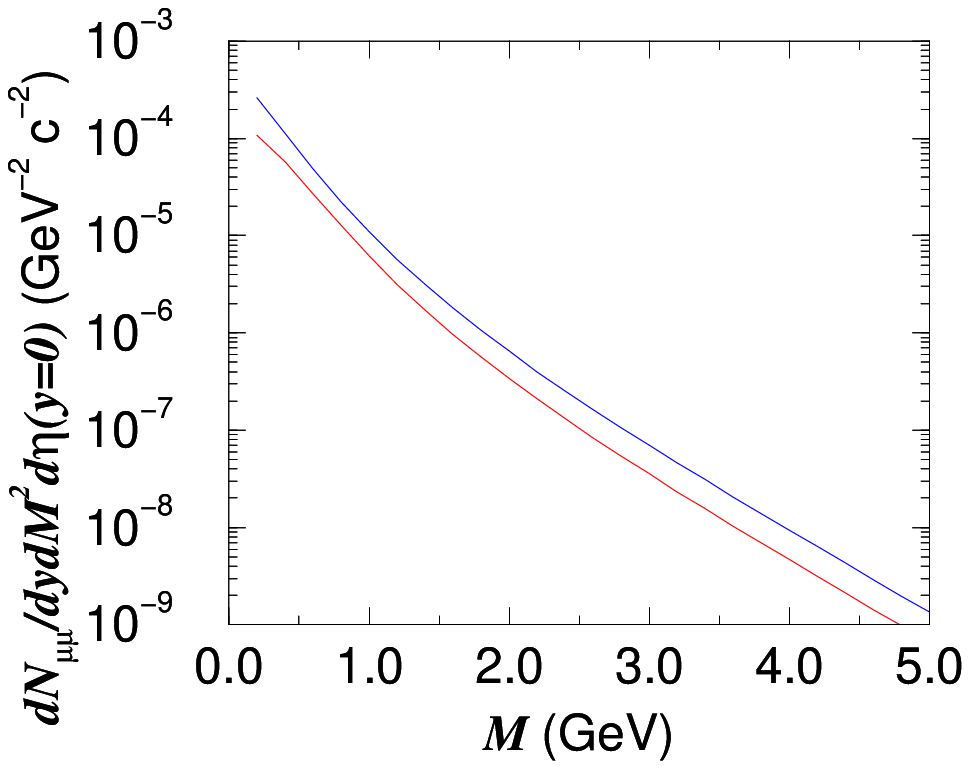,width=2.5in}}
}}
\caption{Same as in \fref{f:dilep} but at RHIC. Again,
emission from ISIPP (lower) is below that from a standard
parton plasma (upper).}
\label{f:dilep2}
\efig

Since the leading order contribution is only subjected 
to indirect effect of the coupling, we also work out 
the next-to-leading order contributions, from non-interference 
graphs only, to get a glimpse of higher orders. 
We have to mention that the following is not a consistent 
calculation of higher order contributions because the 
time evolution of the plasma was done with interactions at 
leading order \cite{wong3,wong2}
and so the following results should be viewed 
as an explorative study. The non-interference Feynman graphs 
at the next-to-leading order are similar to those of the photon 
production except the photon is now off-shell and timelike. 
Again as in real photon production, we have Compton and 
annihilation contributions. The production rate is
\bea \frac{d^4 N^{(\a_s)}_{\m^+\m^-}}{d^4 x} \fx &=& \fx 
     \int \intkso \intkst \intksth \intlso                  \nonum 
     \fx & \times & \fx 
         \; \intlst  (2\p)^4 \d^{(4)} (k_1+k_2-k_3-l_1-l_2) \nonum     
     \fx & \times & \fx  \bigg \{
     2 f_g (k_1,\t) f_q (k_2,\t) (1-f_q(k_3,\t))            \nonum
     \fx & & \fx \;\;\; \times |\cm_{gq\lra q\m^+\m^-}|^2   
                                                            \nonum
     \fx & & \fx \;\;
    +  f_q (k_1,\t) f_q (k_2,\t) (1+f_g(k_3,\t))            \nonum
     \fx & & \fx \;\;\; \times |\cm_{q\bar q\lra g\m^+\m^-}|^2             
     \bigg \}   \; .                                             
\eea

This can be rewritten as
\bea \frac{d^8 N^{(\a_s)}_{\m^+\m^-}}{d^4 x d^4 q} \fx &=& \fx 
      \int \intkso \intkst \intksth                         \nonum 
     \fx & \times & \fx \d^{(4)} (k_1+k_2-k_3-q) 
           \frac{2 q^0 \Gamma_{\g^* \lra \m^+\m^-}(M)}{M^4} \nonum     
     \fx & \times & \fx  \bigg \{
     2 f_g (k_1,\t) f_q (k_2,\t) (1-f_q(k_3,\t))            \nonum
     \fx & & \fx \;\;\; \times |\cm_{gq\lra q\g^*}|^2   
                                                            \nonum
     \fx & & \fx \;\;
    +  f_q (k_1,\t) f_q (k_2,\t) (1+f_g(k_3,\t))            \nonum
     \fx & & \fx \;\;\; \times |\cm_{q\bar q\lra g\g^*}|^2             
     \bigg \}   
\eea
in terms of the decay width of a timelike virtual photon 
into a dilepton pair, which is given by
\bea \Gamma_{\g^* \lra \m^+\m^-}(M) \fx &=& \fx \frac{1}{2q^0} 
     \int \intlso \intlst                                   \nonum
     \fx & \times & \fx (2\p)^4 \d^{(4)} (q-l_1-l_2) 
     | \cm_{\g^* \lra \m^+\m^-} |^2    \; .                 \nonum
\eea
The sum over final states and averaged over initial state
matrix element squared is 
$|\cm_{\g^* \lra \m^+\m^-} |^2=2^4\p \a M^2/3$ and 
so the decay width is $\Gamma_{\g^* \lra \m^+\m^-}(M)=\a M^2/3 q^0$.

The other matrix element squared for virtual photon production 
\cite{field} via Compton scattering, including again colour, 
spin and infrared screening, is
\bea |\cm_{gq\lra g\g^*} |^2 \fx &=& \fx 
     \sum_q e^2_q 2^9 \p^2 \a \a_s                          
     \bigg \{-\frac{t}{s+m_q^2}-\frac{s}{t-m_q^2}           \nonum
     \fx & & \fx
             +2 M^2 \Big ( \;\;
                     \frac{1}{t-m_q^2}+\frac{1}{s+m_q^2}    \nonum   
     \fx & & \fx \mbox{\hskip 1.0cm} 
             -  \frac{M^2}{(s+m_q^2)(t-m_q^2)} 
                    \Big ) 
     \bigg \}       
\eea
and that from annihilation is
\bea |\cm_{q\bar q\lra g\g^*} |^2 \fx &=& \fx 
     \sum_q e^2_q 2^9 \p^2 \a \a_s                          
     \bigg \{\frac{t}{u-m_q^2}+\frac{u}{t-m_q^2}            \nonum
     \fx & & \fx
             -2 M^2 \Big ( \;\;
                     \frac{1}{t-m_q^2}+\frac{1}{u-m_q^2}    \nonum   
     \fx & & \fx \mbox{\hskip 1.0cm} 
             -  \frac{M^2}{(u-m_q^2)(t-m_q^2)} 
                    \Big ) 
     \bigg \}       \; .
\eea
There is an additional infrared divergence hidden in $f_g(k_3)$ 
in the annihilation contribution when very soft gluon is
emitted with the photon, we cut this off by requiring
$k_3 \ge m_g$, the gluon mass in the medium. This problem
is not present in real photon production because there $k_3$ 
can only reach zero when $s=0$ when the modulus squared of the 
matrix element vanishes. This latter in the present virtual photon
case does not vanish. We use $m_g^2 = \frac{1}{3} m_D^2$, 
the relation between the gluon mass and the Debye screening 
mass in equilibrium, and $m_D^2$ is calculated similarly 
from the time-dependent quark and gluon distribution 
\cite{wong1,wong2}.

\bfig
\centerline{
\hbox{
{\psfig{figure=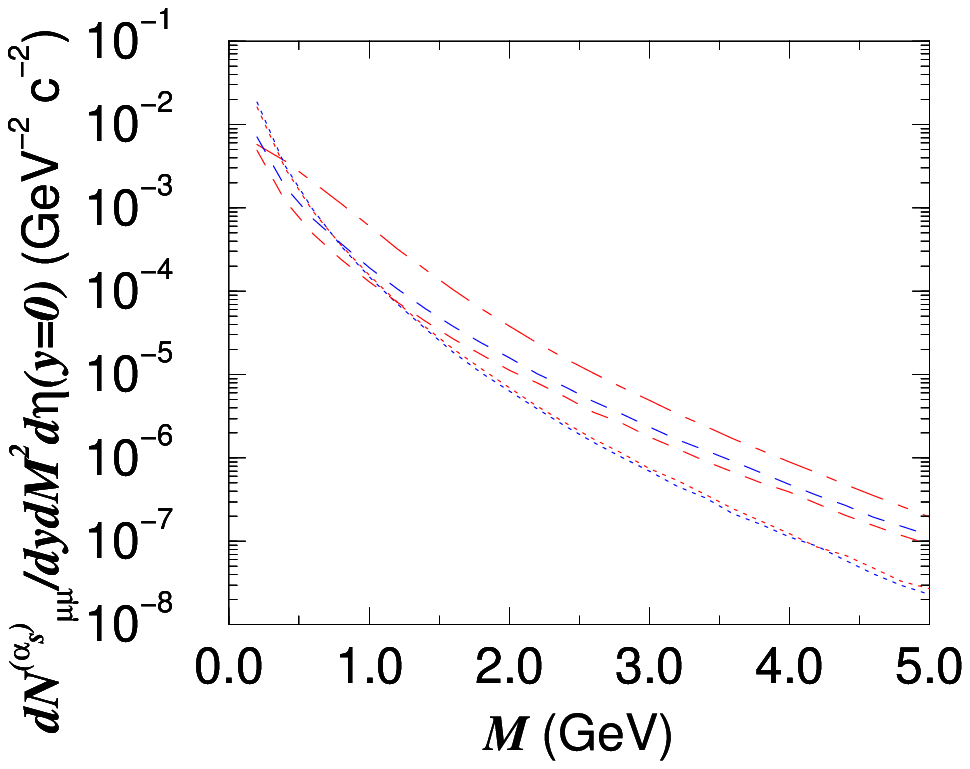,width=2.5in}}
}}
\caption{Dilepton emission at next-to-leading order at
LHC from ISIPP and from the standard parton plasma.
Timelike virtual photon emissions from 
quark-antiquark annihilation (dashed) dominate over those 
from Compton scattering (dotted) at higher values and in fact
most values of $M$ because of the combined effect of 
interference and quantum statistics. The two sets of curves 
from ISIPP and from the standard plasma lie almost on top 
of each other with the annihilation contribution from the
standard plasma slightly higher due to the smaller cutoff
of the gluon mass in the medium, and so are, similar to 
real photon production at leading order, essentially 
the same for both scenarios. Emission at leading order 
from ISIPP (dot-dashed) is plotted again for comparison.} 
\label{f:dilep_ho}
\efig

\bfig
\centerline{
\hbox{
{\psfig{figure=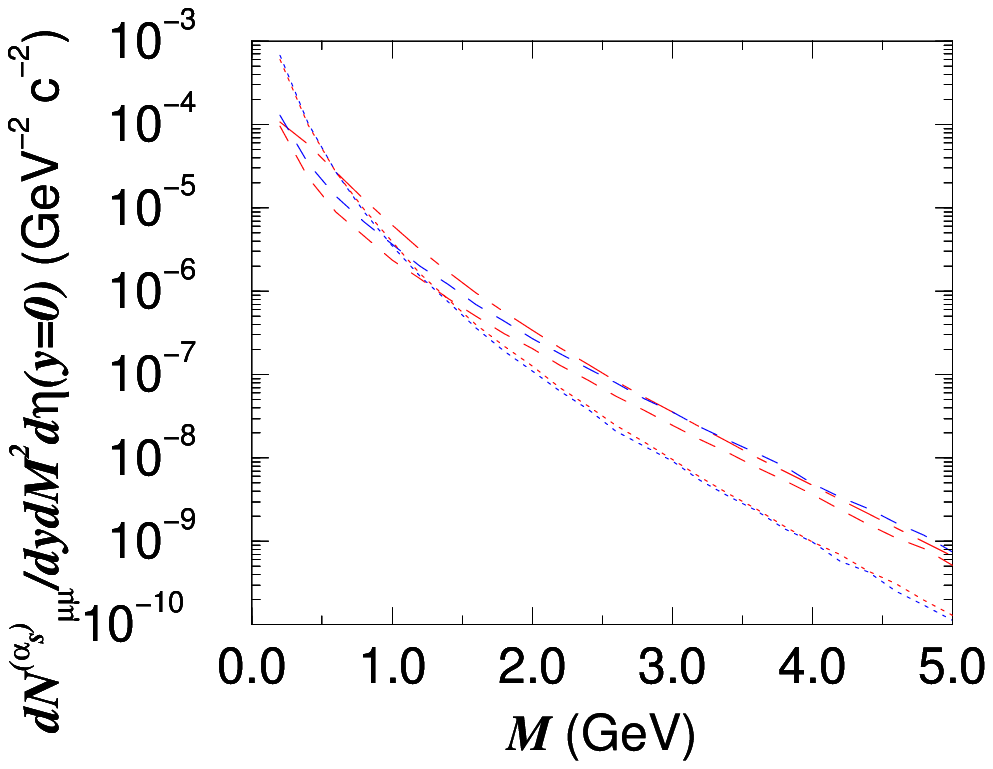,width=2.5in}}
}}
\caption{Same as in \fref{f:dilep_ho} but at RHIC.
In this case, the annihilation contribution 
(lower dashed) is much closer to the 
leading order contribution (dot-dashed) from ISIPP at 
larger $M$.}
\label{f:dilep2_ho}
\efig

The next-to-leading order results are plotted in 
\fref{f:dilep_ho} at LHC and in \fref{f:dilep2_ho} at RHIC.
Like photon production at leading order, negative and positive 
direct and indirect effects largely cancel out each other 
so that the invariant mass distribution from ISIPP and from 
the standard parton plasma are very similar. The small 
decrease of the more dominant annihilation contribution
from ISIPP, the lower of the two dashed curves in each
figure, is due to the gluon mass cutoff used, which  
is larger in ISIPP. In the present case, annihilation
contribution dominates at large $M$ because of the interference
effect in the matrix elements for timelike virtual photon
emission. This is the more important, the larger the value
of $M$. This reduces Compton scattering but increases
annihilation contribution. This is however not a medium effect.
What is the result of medium effect and has to do with the
increasing interaction strength is the difference of the
production rate between the leading and next-to-leading order.
If one examines, for example \fref{f:dilep2} and \fref{f:dilep2_ho}, 
the leading order and the next-to-leading 
order are closer to each other in ISIPP than in the standard
parton plasma, so higher orders are more comparable to the
leading order especially at RHIC energies. 
Therefore in general in ISIPP, higher orders are more important 
for dilepton production in heavy ion collisions as a result 
of the increasing coupling effect.

\section{Summary and Outlook}
\label{sec:end}

We have examined the effects of the coupling and 
out-of-equilibrium environment had on three types of
particle production in the new ISIPP scenario. We found
that although at first sight there seemed to be 
an unavoidable reduction in the open charm yield, the
concentration of the contribution from gluon in the 
early times prevented a significant diminution in the
yield. As such, the usefulness of open charm as a probe
of early parton densities is not affected. For photon
production, a non-equilibrium effect not shown previously
which came about as a result of quark-antiquark and 
gluon existing as a mixed fluid with different effective
temperatures and with different cooling rates, brought
the quark and antiquark annihilation contribution to photon
production, in the higher $p_T$ region, to above that
from Compton scattering. Whereas in a thermally 
equilibrated plasma, Compton scattering would dominate
in the same region. For dilepton production, something similar 
happened but this time, it was due to quantum interference 
effect. The more interesting medium effect we found by
an explorative study into higher order contributions in this
case was that ISIPP rendered the leading and next-to-leading
order contributions to be comparable with each other especially 
at RHIC. Thus higher orders seem to have greater significance 
in ISIPP than in the standard parton plasma. This could be 
viewed as a possible representative result of how higher 
orders could affect non-hard type processes such as 
the electromagnetic productions considered
here. But before we can study particle production more 
closely, higher orders have first to be included in
the time evolution of the system.

In this work, we restricted our considerations on the 
effects of ISIPP on particle productions to the deconfined 
phase. Apparently, there is a lack of enhancement that
one would hope for. However, bearing in mind that the
ratio of hadronic and partonic signals is more crucial
for the search for the quark-gluon plasma, a reduction
in hadronic signals would be just as good. It must be
mentioned that the effects of ISIPP extend well beyond 
the parton phase because of the effects on the entropy 
during the equilibration of the parton plasma. 
As shown in \cite{wong3}, entropy would be 
reduced more significantly in the parton plasma,
the larger is the coupling. So entropy reduction in ISIPP would
reduce the duration of the mixed phase if there is a
first order phase transition. The signals from this phase
will be affected in a certain way. One can imagine
the increasing coupling effects of the direct and indirect
type will continue to be effective for signals of partonic 
origin. But for signals of hadronic origin, they will only be
subjected to indirect effects. These, when applied to the
hadronic part of the mixed phase and the subsequent hadron
phase, apart from the reduced duration of the former already 
mentioned, in view of the entropy reduction, will lead to a
lowered final hadron multiplicity and hence hadron gas density. 
The latter would means weaker hadronic signals. Since direct 
effect tends to be stronger than indirect ones, if higher 
orders are taken into account, then signals
of partonic origin, but not those of hard processes, 
will at least remain the same if not actually be enhanced,
but those of hadronic origin will be suppressed. 
Therefore the effects of the interaction strength of the 
parton plasma could shift the balance of certain signals, 
for example electromagnetic ones, emitted from relativistic heavy 
ion collisions experiments in favour of those of partonic 
origin and thus the search for the quark-gluon plasma would 
be made easier than before. 

While we were finishing this paper, we received 
two papers \cite{let&etal1,let&etal2}, in which the authors 
also pointed out quite rightly and discussed the importance 
of running both in the strong coupling and in the quark mass 
in thermal flavour production. It seems that one should also 
consider this in the initial as well as in the early minijet 
gluon conversion stage especially for charm for the purpose
of comparison. As we discussed here, the running coupling
effects are more general and wide ranging and should be
taken into account not only for flavour production in the
quark-gluon plasma.

\section*{Acknowledgements}

The author would like to thank K. Redlich for useful discussion
on photon production.

\vfill\break

\null
\vfill\break

\section*{Figure Captions}

\begin{itemize}

\item[\fref{f:charm}]{Comparison of charm production in a parton plasma 
produced at LHC energies, which is time evolved with an evolving 
coupling $\a_s^v$, with the one time evolved with a 
fixed $\a_s=0.3$. For the production itself, we used 
$\a_s=0.3$ in all cases because this is a hard process. 
Dotted and dashed lines are for gluon conversion 
and quark-antiquark annihilation contribution respectively. 
Solid lines are the sum total. Lines from ISIPP tend to 
lie slightly below the corresponding lines from the fixed
$\a_s =0.3$ evolved plasma. It is more clearly so for 
annihilation contribution.}

\item[\fref{f:charm2}]{Same as in \fref{f:charm} but for a parton plasma 
produced at RHIC energies.}

\item[\fref{f:phot}]{Photon production from the parton plasma at LHC.
The solid lines are the total sum of the emission from Compton
scattering (dotted) and quark-antiquark annihilation 
(dashed). At large $p_T$, quark-antiquark annihilation 
is the dominant contribution because of the fact that
quarks and gluons are not in equilibrium with respect to 
each other and are therefore at different effective 
temperatures and of quantum statistical effect to a
lesser extent. This contribution from the standard parton 
plasma is slightly above that from ISIPP at large $p_T$.}

\item[\fref{f:phot2}]{Same as \fref{f:phot} but at RHIC. 
In this case, however,
emissions from Compton scattering remain above those from 
quark-antiquark annihilation up to $p_T=$5.0 GeV because of the 
much lower quark to gluon density ratio at RHIC. The point 
where the emission from the latter begin to dominate over the 
former is at higher $p_T$ beyond 5.0 GeV.}

\item[\fref{f:dilep}]{Comparison of dilepton emission from an 
ordinary parton plasma (upper solid line) with ISIPP (lower solid line) 
at LHC. Even though the fermion densities are enhanced in 
ISIPP, the shortening of the duration of the plasma in the 
parton phase is the more important of the two effects. 
So the emission from ISIPP is reduced.}

\item[\fref{f:dilep2}]{Same as in \fref{f:dilep} but at RHIC. Again,
emission from ISIPP (lower) is below that from a standard
parton plasma (upper).}

\item[\fref{f:dilep_ho}]{Dilepton emission at next-to-leading 
order at LHC from ISIPP and from the standard parton plasma.
Timelike virtual photon emissions from 
quark-antiquark annihilation (dashed) dominate over those 
from Compton scattering (dotted) at higher values and in fact
most values of $M$ because of the combined effect of 
interference and quantum statistics. The two sets of curves 
from ISIPP and from the standard plasma lie almost on top 
of each other with the annihilation contribution from the
standard plasma slightly higher due to the smaller cutoff
of the gluon mass in the medium, and so are, similar to 
real photon production at leading order, essentially 
the same for both scenarios. Emission at leading order 
from ISIPP (dot-dashed) is plotted again for comparison.} 

\item[\fref{f:dilep2_ho}]{Same as in \fref{f:dilep_ho} but 
at RHIC. In this case, the annihilation 
contribution (lower dashed) is much closer to the 
leading order contribution (dot-dashed) from ISIPP at 
larger $M$.}

\end{itemize}


\begin{thebibliography}{99}

\bibitem{mclerr&toi}L.D. McLerran and T. Toimela, 
\Journal{\PRD}{31}{545}{1985}.

\bibitem{kaj&etal}K. Kajantie, J. Kapusta, L. McLerran and A. Mekjian,
\Journal{\PRD}{34}{2746}{1986}.

\bibitem{braa&etal}E. Braaten, R.D. Pisarski and C.P. Yuan,
\Journal{\PRL}{64}{2242}{1990}.

\bibitem{kap&etal}J. Kapusta, P. Lichard and D. Seibert,
\Journal{\PRD}{44}{2774}{1991}.

\bibitem{weld}H.A. Weldon, \Journal{\PRL}{66}{293}{1991}.

\bibitem{baier&etal1}R. Baier, H. Nakkagawa, A. Ni\'egawa
and K. Redlich, \Journal{\ZPC}{53}{433}{1992}.

\bibitem{wong4}S.M.H. Wong, \Journal{\ZPC}{53}{465}{1992}.

\bibitem{chak&etal}S. Chakrabarty, J. Alam, D.K. Srivastava and
B. Sinha, \Journal{\PRD}{46}{3802}{1992}. 

\bibitem{geig&etal}K. Geiger and J.I. Kapusta, 
\Journal{\PRL}{70}{1920}{1993}.

\bibitem{shury&xion2}E.V. Shuryak and L. Xiong, 
\Journal{\PRL}{70}{2241}{1993}.

\bibitem{trax&etal1}C.T. Traxler, H. Vija and M.H. Thoma,
\Journal{\PLB}{346}{329}{1995}.

\bibitem{trax&thoma}C.T. Traxler and M.H. Thoma,
\Journal{\PRC}{53}{1348}{1996}.

\bibitem{pal&etal}D. Pal, P.K. Roy, S. Sarkar, D.K. Srivastava,
and B. Sinha, \Journal{\PRC}{55}{1467}{1997}.

\bibitem{baier&etal2}R. Baier, M. Dirks and K. Redlich,
\Journal{\PRD}{55}{4344}{1997}.

\bibitem{cley&etal}J. Cleymans, R. Redlich and D.K. Srivastava,
preprint nucl-th/9702004. 

\bibitem{mull&etal}B. M\"uller, M.G. Mustafa, D.K. Srivastava,
\Journal{\PRC}{56}{1064}{1997}.

\bibitem{muell&wang}B. M\"uller and X.N. Wang, 
\Journal{\PRC}{51}{3326}{1995}. 

\bibitem{geig2}K. Geiger, \Journal{\PRD}{48}{4129}{1993}.

\bibitem{lin&gyul}Z. Lin and M. Gyulassy, 
\Journal{\PRC}{51}{2177}{1995}. 

\bibitem{lev&etal}P. L\'evai, B. M\"uller and X.N. Wang, 
\Journal{\PRC}{51}{3326}{1995}.

\bibitem{eskola}K.J. Eskola, \Journal{\NPB}{400}{240}{1993}.

\bibitem{bern&etal}W. Bernreuther, A. Brandenburg and P. Uwer,
\Journal{\PRL}{79}{189}{1997}.

\bibitem{marti&etal}S. Mart\'{\i} i Garc\'{\i}a, J. Fuster and 
S. Cabrera, talk presented at QCD'97, Montpellier, France, July 97,
hep-ex/9708030.

\bibitem{bil&etal}M. Bilenky, G. Rodrigo and A. Santamaria,
\Journal{\PRL}{79}{193}{1997}.

\bibitem{wong3}S.M.H. Wong, \Journal{\PRC}{56}{1075}{1997}.

\bibitem{shury}E.V. Shuryak, \Journal{\PRL}{68}{3270}{1992}.

\bibitem{wong1}S.M.H. Wong, \Journal{\NPA}{607}{442}{1996}.

\bibitem{wong2}S.M.H. Wong, \Journal{\PRC}{54}{2588}{1996}.

\bibitem{comb}B.L. Combridge, \Journal{\NPB}{151}{429}{1979}.

\bibitem{gyu1}M. Gyulassy and X.N. Wang, \Journal{\PRD}{44}{3501}{1991}.

\bibitem{gyu2}M. Gyulassy, M. Pl\"umer, M. Thoma and X.N. Wang, 
\Journal{\NPA}{538}{37c}{1992}.

\bibitem{gyu3}M. Gyulassy and X.N. Wang, \Journal{\NPA}{544}{559c}{1992}.

\bibitem{huang&etal}Z. Huang, H.J. Lu and I. Sarcevic, 
preprint AZPH-TH/97-07, hep-ph/9705250.

\bibitem{gupta}S. Gupta, \Journal{\PLB}{248}{453}{1990}.

\bibitem{field}R.D. Field, {\it Applications of Perturbative QCD},
Frontiers in Physics Vol. 77 (Addison-Wesley, Redwood City, CA, 1986).

\bibitem{let&etal1}J. Letessier, J. Rafelski and A. Tounsi,
\Journal{\PLB}{389}{586}{1996}.

\bibitem{let&etal2}J. Letessier, J. Rafelski and A. Tounsi,
in {\it Proceedings of ICHEP'96}, edited by Z. Ajduk and 
K. Wrobleski. (World Scientific, Singapore, 1997), p. 971. 

\end{thebibliography}
\end{document}